\documentclass[a4paper,10pt,twoside]{cpc-hepnp}
\usepackage{CJK,upgreek,fancyhdr}
\usepackage{multicol}
\usepackage{graphicx}
\usepackage{booktabs}
\usepackage{amssymb,bm,mathrsfs,bbm,amscd}
\usepackage[tbtags]{amsmath}
\usepackage{lastpage}
\usepackage{color}

\begin{document}
\begin{CJK*}{GBK}{song}

\fancyhead[c]{\small Submitted to Chinese Physics C}
\fancyfoot[C]{\small 010201-\thepage}


\title{Triple $\alpha$ Resonances in the Decay of Hot Nuclear Systems\thanks{This work was supported by the National Natural Science Foundation of China (No. 1176014, 11605097, 11421505,11865010),the US Department of Energy under Grant No. DE-FG02-93ER40773, NNSA DE-NA0003841 (CENTAUR) and the Robert A. Welch Foundation under Grant No. A-1266.
This work was also supported by the Chinese Academy of Sciences (CAS) President's International Fellowship Initiative (No. 2015VWA070),
Strategic Priority Research Program of the Chinese Academy of Sciences (No. XDB16 and XDPB09), Program for Young Talents of Science and Technology in
Universities of Inner Mongolia Autonomous Region (NJYT-18-B21) and Doctoral Scientific Research Foundation of Inner Mongolia
University for Nationalities (No. BS365 and BS400), The Fundamental Research Funds for the Central Universities (No. GK201903022), Natural Science Foundation of Inner Mongolia(No. 2018MS01009).}}

\author{%
      S. Zhang$^{1)}$\email{zsylt@imun.edu.cn}%
\quad J.C. Wang$^{1}$
\quad A. Bonasera$^{2,3}$
\quad M. Huang$^{1}$
\quad H. Zheng$^{4}$
\\ G. Zhang$^{5,6}$
\quad Z. Kohley$^{2,7}$
\quad Y.G. Ma$^{5,6}$
\quad S.J. Yennello$^{2,7}$
}
\maketitle

\address{%
$^1$College of Physics and Electronics information, Inner Mongolia University for Nationalities, Tongliao, 028000, China\\
$^2$Cyclotron Institute, Texas A$\&$M University, College Station, Texas 77843, USA\\
$^3$Laboratori Nazionali del Sud, INFN, via Santa Sofia, 62, 95123 Catania, Italy\\
$^4$School of Physics and Information Technology, Shaanxi Normal University, Xi'an 710119, China\\
$^5$Shanghai Institute Applied Physics, Chinese Academy of Sciences, Shanghai 201800, China\\
$^6$Key Laboratory of Nuclear Physics and Ion-beam Application (MOE), Institute of Modern Physics, Fudan University, Shanghai 200433, China\\
$^7$Chemistry Department, Texas A\&M University, College Station, Texas 77843, USA\\
}

\begin{abstract}
The Efimov (Thomas) trimers in excited $^{12}$C nuclei, for which no observation exists yet, are discussed by means of analyzing the experimental data of $^{70(64)}$Zn($^{64}$Ni) + $^{70(64)}$Zn($^{64}$Ni) reactions at beam energy of E/A=35 MeV/nucleon. In heavy ion collisions, the $\alpha$s interact with each other and can form complex systems such as $^{8}$Be and $^{12}$C. For the 3$\alpha$ systems, multi resonance processes give rise to excited levels of $^{12}$C. The interaction between any two of the 3$\alpha$ particles provides events with one, two or three $^{8}$Be. Their interfering levels are clearly seen in the minimum relative energy distributions. Events of three couple $\alpha$ relative energies consistent with the ground state of $^{8}$Be are observed with the decrease of the instrumental error at the reconstructed 7.458 MeV excitation energy of $^{12}$C, which was suggested as the Efimov (Thomas) state.
\end{abstract}



\footnotetext[0]{\hspace*{-3mm}\raisebox{0.3ex}{$\scriptstyle\copyright$}2013
Chinese Physical Society and the Institute of High Energy Physics
of the Chinese Academy of Sciences and the Institute
of Modern Physics of the Chinese Academy of Sciences and IOP Publishing Ltd}%

\begin{multicols}{2}

\section{Introduction}
In 1969, Vitaly Efimov, following a work by Thomas(1935) \cite{0Thomas}, first predicted a puzzling quantum-mechanical effect, when a resonant pairwise interaction gives rise to an infinite number of three-body loosely bound states even though each particle pair is unable to bind \cite{1Efimov:1970zz,8efimov:nature09}. These properties are universal and independent of the details of the short-range interaction when the two-body scattering length `a' is much larger than the range of the interaction potential `r$_0$'. The existence of resonant two-body forces is the basic condition for the Efimov effect. Although there has been an extensive search in many different physical systems including atoms, molecules and nuclei, the experimental confirmation of existence of Efimov states has proved to be challenging especially for nuclei \cite{0Thomas,1Efimov:1970zz,8efimov:nature09, 2Braaten2015, 3Tumino:2015jaa, 4Kraemer2006, 5Zaccanti2009, 6Huang2014, 7Gattobigio2014}.
Recently, Tumino {\it et al.} reported about the discovery of the existence of triple-alpha resonances, very close to the Efimov scenario, by studying $^{6}$Li+$^{6}$Li$\rightarrow$3$\alpha$ reactions at low beam energy and using hyperspherical formalism.
A geometrical interpretation of these mechanisms \cite{HuaarXiv2018} suggests that the Thomas state corresponds to three equal energies, while a sequential decay mechanism ($^{12}$C$\rightarrow$$^{8}$Be+$\alpha$$\rightarrow$3$\alpha$) might correspond to Efimov states \cite{1Efimov:1970zz}. This prescription refers mainly to $^{12}$C levels in the vicinity of the breakup threshold of three $\alpha$-particles or $\alpha$+$^{8}$Be, taking into account the Coulomb force among $\alpha$ particles which destroys the 1/R$^2$ (R is the hyperradius) scaling at large distance where Coulomb force is dominant \cite{1Efimov:1970zz}.
This is surely relevant with stellar processes, where $^{12}$C nucleus is formed and it might occur inside a dense star or on its surface, thus in different density and temperature conditions. A way to simulate some stellar conditions is to collide two heavy ions at beam energies near the Fermi energy.
In this work, we present the possible signature of Efimov (Thomas) states at reconstructed 7.458 MeV excitation energy of $^{12}$C from the reactions $^{70(64)}$Zn($^{64}$Ni) + $^{70(64)}$Zn($^{64}$Ni) at beam energy of E/A=35 MeV/nucleon \cite{SuyalatuarXiv2018}.

\section{Experiment}
The experiment was performed at the Cyclotron Institute, Texas A$\&$M University. $^{64}$Zn, $^{70}$Zn, and $^{64}$Ni beams at 35 MeV/nucleon from the K-500 superconducting cyclotron were used to respectively irradiate self-supporting $^{64}$Zn, $^{70}$Zn, and $^{64}$Ni targets. The 4$\pi$ NIMROD-ISiS setup \cite{9Wuenschel2009, 10Schmitt1995} was used to collect charged particles and free neutrons produced in the reactions. A detailed description of the experiment can be found in Refs. \cite{11Kholey:2010phd, 12Kohley2011, 13Kohley2012}.

\begin{center}
\includegraphics[scale=0.35]{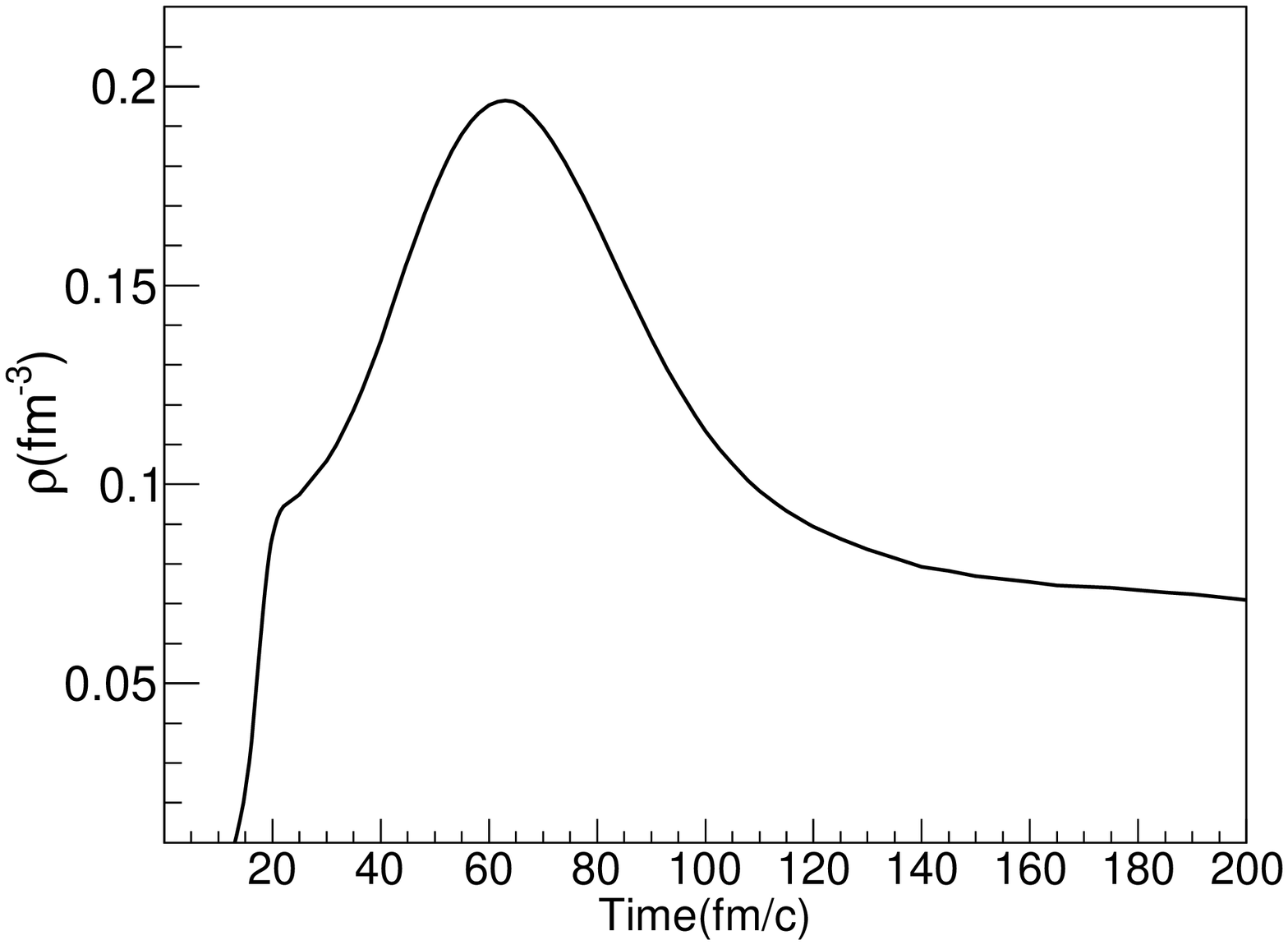}
\figcaption{\label{fig1_density} The time evolution of the average density in the central region in $^{70}$Zn + $^{70}$Zn at 35 MeV/nucleon.
}
\end{center}

When two heavy ions at 35 MeV/nucleon collide, the excitation energy deposited in the system is large enough for the system to get gently compressed at the beginning and then it expands and enters an instability region, the spinodal region, similar to the liquid-gas (LG) phase transition \cite{14Giuliani2014, 15Bonasera:1994pr, mayg:1999prl,mayg:2005prc,mayg:2018ppnp}. Fig. \ref{fig1_density} shows the time evolution of the average density in the central region $[-3,3]^{3} fm^{3}$ at incident energy of 35 MeV/nucleon in collisions of $^{70}$Zn + $^{70}$Zn with the Constrained Molecular Dynamics approach (CoMD) \cite{14Giuliani2014}. The average density increases in the compression stage and decreases in the expansion stage. The maximum average density reaches around 60 fm/c when the initial distance between projectile and target nuclei is set to 15 fm.
In such conditions, fragments of different sizes are formed and later can be detected. The NIMROD detector used in this experiment can distinguish charge numbers from 1 to 30 and masses up to 50 \cite{11Kholey:2010phd}. A typical result is plotted in Fig. \ref{fig2_yield} \cite{11Kholey:2010phd} together with the CoMD results \cite{14Giuliani2014}, showing a satisfactory agreement to the data. In order to test if some fragments are formed in excited states, an evaporation model, Gemini \cite{14Giuliani2014,mayg:2002prc,mayg:2018nst,mayg:2006cpc,gemini:2018cpc} is applied. The reaction was followed up to a maximum time t$_{max}$  in the CoMD model. Within the same model, the excitation energy of each fragment formed at t$_{max}$ is obtained and fed into the Gemini model, which gives the final de-excited fragments. As can be seen from the figure, the effect of secondary evaporation is negligible after t$_{max} >$600 fm/c. The abundance of $^{12}$C fragments are about two orders of magnitude less than those of proton and $\alpha$-particles. These ions survive the violence of the collision while other $^{12}$C might be in one of the excited states and decays before reaching the detector or collides with other fragments and gets destroyed. Our technique is tailored to select the $^{12}$C$\rightarrow$3$\alpha$ decay channel among all the possibilities.

\begin{center}
\includegraphics[scale=0.45]{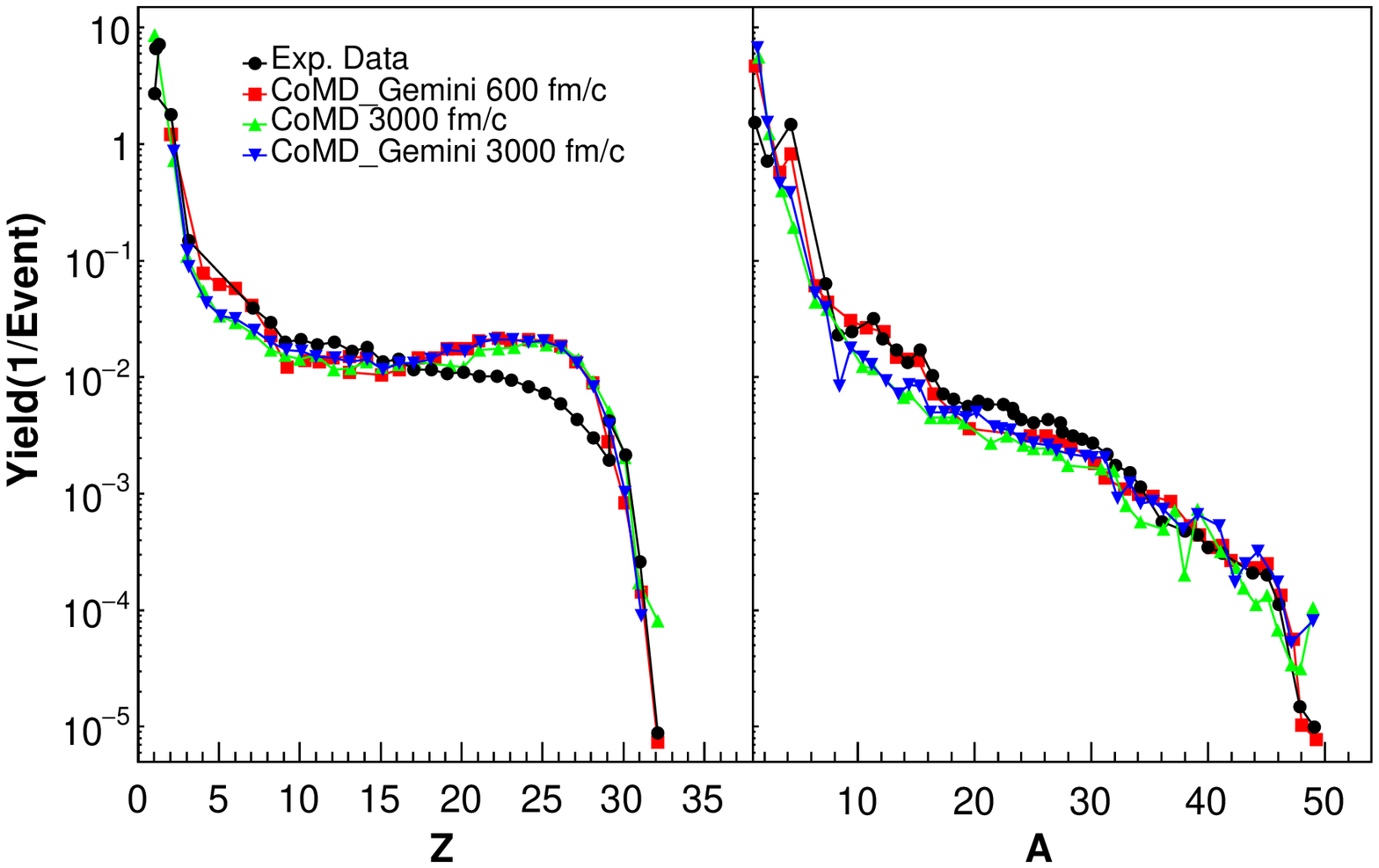}
\figcaption{\label{fig2_yield} (color online) The charge (Z) and mass (A) distributions from the $^{70}$Zn+$^{70}$Zn system are shown for the filtered CoMD simulation in comparison to the experimental data. The results have been normalized by the total number of events \cite{11Kholey:2010phd}.
}
\end{center}

In the experiments, it is straightforward to select all the events where one or more $\alpha$ particles are detected. In Fig. \ref{fig3_mult}, we plot the $\alpha$ particle multiplicity distribution for the three colliding systems considered. The total number of events is $\sim 2.7 \times 10^8$ and we have observed events where at least 15 $\alpha$ are produced. In Refs. \cite{Schmidt:2016lpt, Mabiala:2016gpt, Marini:2015zwa}, an analysis was performed for events as in Fig. \ref{fig3_mult} in terms of Boson-Fermion mixtures, i.e. including all fragments as reported in Fig. \ref{fig2_yield}, which can give also a signature of Bose Einstein Condensation (BEC)  \cite{SchuckPRL2001, SchuckPRL2008}. Temperature, density and excitation energy are recovered using different approaches \cite{14Giuliani2014} with most of the events in the high excitation energy region up to about 8 MeV/nucleon. We notice that most of the novel techniques discussed in this work might be easily generalized to cases where the $\alpha$ multiplicity is larger than 3. We plan to discuss this in future papers. More conventional analysis based on Dalitz plots
\cite{Freer:1994zz,Raduta:2011yz,Kirsebom:2012zza,Manfredi:2012zz,Rana:2013hka,Itoh:2014mwa,DellAquila:2017ppe,Smith:2017jub} cannot be easily generalized when the $\alpha$-multiplicity is larger than 3.

For the purpose of our work, we further selected all the events with only three $\alpha$ particles detected.
It is important to stress that multiple $\alpha$ revealed are accepted if they hit different detectors, i.e. all possible double hits in an event are excluded. Furthermore, in the present analysis, a random position on the surface of the detector was assigned to each $\alpha$. This limits the precision of $\alpha$-$\alpha$ correlations especially when their relative energies or momenta are very small. A critical comparison of different methods to assign the hitting position on the detector will be discussed in a future work, here it is sufficient to say that the results discussed are independent on the different methods.
In these cases, the total number of events reduced to $\sim 4.5 \times 10^7$. From the above discussion, it is clear that if
only three $\alpha$s are in an event, other fragments must be presented and the sum of all the fragment masses is up to 140 (maximum) including
 the three $\alpha$s. This is a rich environment depending on the proximity of different fragments to the $\alpha$, $^{8}$Be or $^{12}$C ions,
 the properties and shell structure of the fragments might be modified. In particular, short living states of $^{12}$C or $^{8}$Be might be modified
 by the presence of other nearby fragments. On the other hand, long living states such as the Hoyle state of $^{12}$C might not be influenced at all
 since its lifetime is much longer than the reaction time. Of course, in such `soup', $\alpha$-particles might come from the decay of $^{12}$C or $^{8}$Be and
 from different excited fragments, or directly produced during the reaction, thus it is crucial to implement different methods to distinguish among different decay channels.

\begin{center}
\includegraphics[scale=0.3]{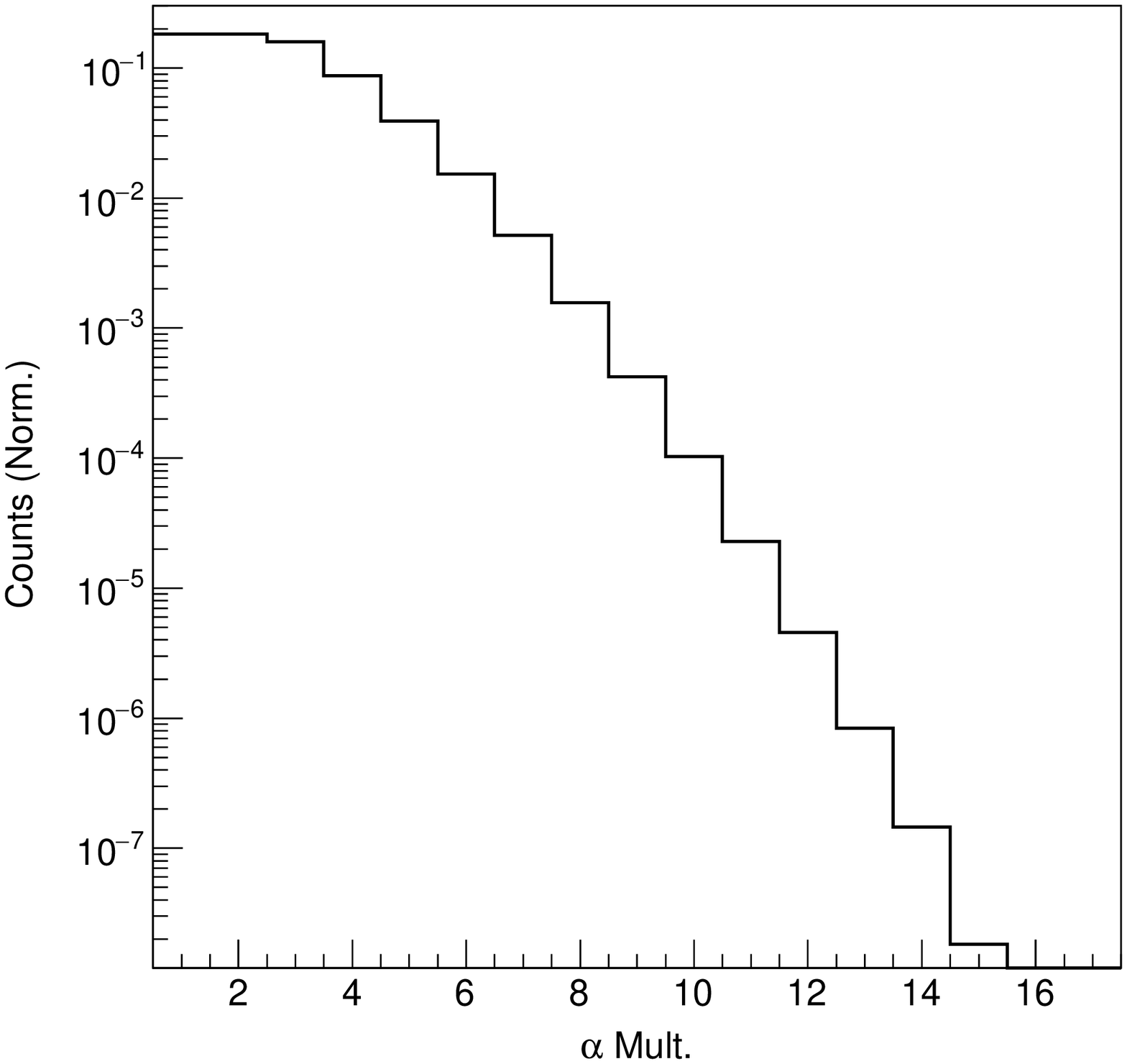}
\figcaption{\label{fig3_mult} $\alpha$ multiplicity distribution for the $^{70(64)}$Zn($^{64}$Ni) + $^{70(64)}$Zn($^{64}$Ni) at 35 MeV/nucleon from the NIMROD detector.
}
\end{center}

In order to distinguish different decay channels, the kinetic energy of the $\alpha$ particles must be measured in a good precision. The kinetic energy distribution from the NIMROD detector for the events with $\alpha$ multiplicity equal to three is given in Fig. \ref{fig4_Ek}. It extends above 100 MeV/nucleon and displays a large yield around
8 MeV/nucleon. Since the kinetic energies are relatively large, the detector is performing its best, and the error estimate (including the instrumental error especially because of the detector granularity, detectors' energy, position, and angle resolution) gives results in less than 1$\%$ of the kinetic energy value. The error becomes larger for smaller kinetic energies and particles whose kinetic energy is below a threshold (about 1 MeV/nucleon) are not detected \cite{11Kholey:2010phd}. Thus it is a clear advantage to use the beam of heavy ions near or above the Fermi energy. Fragments are emitted in the laboratory frame with high kinetic energies (due to the center of mass motion) and can be carefully detected. When we reconstruct $^{8}$Be from $\alpha$-$\alpha$ correlations, the center of mass motion is cancelled out and small relative kinetic energies can be obtained with an estimated error of about 45 keV for the smallest relative kinetic energies. This error is due to the detector granularity as discussed above.

\begin{center}
\includegraphics[scale=0.3]{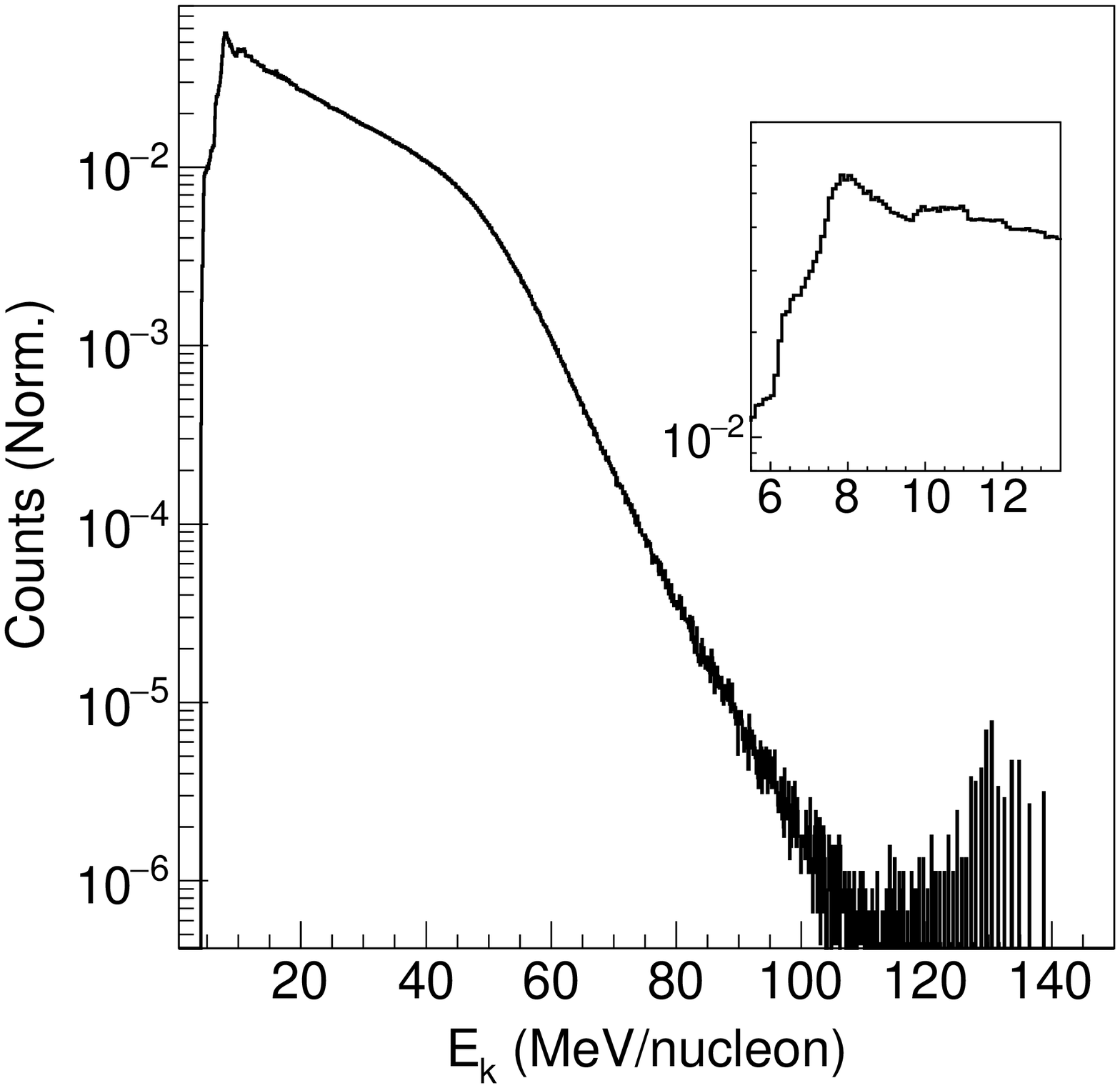}
\figcaption{\label{fig4_Ek} $\alpha$-kinetic energy distribution in the laboratory frame from all the events with $\alpha$-multiplicity equal to three. Inset: zoom of the
lower energy region in this figure.
}
\end{center}

\section{Method}
For the three body system with equal masses, we can define the excitation energy E${^*}$ as:
\begin{equation}
 E^* = \frac{2}{3}\sum^3_{i=1,j>i} E_{ij} - Q
\label{Estar}
\end{equation}
where E$_{ij}$ is the relative kinetic energy of the two particles, Q is the Q-value. Notice that the important ingredients entering Eq.(\ref{Estar}) are the relative kinetic energies; since we have three indistinguishable bosons, we analyze the E$_{ij}$ distribution by cataloging for each event the smallest relative kinetic energy, $E_{ij}^{Min.}$, the middle relative kinetic energy, $E_{ij}^{Mid.}$, and the largest relative kinetic energy, $E_{ij}^{Lar.}$.

In this work, we reconstruct the E${^*}$=7.458 MeV of the $^{12}$C from 3$\alpha$s when the sum of the three E$_{ij}$ is 0.276 MeV (0.092$\times$3 MeV, i.e. 0.092 MeV is the relative energy of 2$\alpha$s corresponding to the ground state decay of $^{8}$Be \cite{HuaPLB2018, SuyalatuarXiv2018}) with the Q-value= -7.275 MeV.
In Fig. \ref{fig5_Eijmin}, the minimum relative kinetic energy distribution is shown. In the top panel, the solid black circles give the distribution obtained from the real events. They show bumps but no real structures. This is due to the fact that in the fragmentation region, some $\alpha$s may come from the decay of $^{8}$Be or $^{12}$C, or  from completely non-correlated processes, for example, the $\alpha$ emission from a heavy fragment. To distinguish the correlation from the non-correlated events, we randomly choose three different $\alpha$s from three different events and build the distribution displayed in Fig. \ref{fig5_Eijmin} (mixing events-red open circles). The total number of real and mixing events are normalized to one, respectively. We fit the highest points of Fig. \ref{fig5_Eijmin} (top) with an exponential function. This allows us to derive the instrumental error $\Delta$E=1/22 MeV=0.045 MeV. By subtracting the fit function from the real events, we obtain the open squares in Fig. \ref{fig5_Eijmin} (top), which can be considered as the real events corrected by the detector acceptance. The ratio (1+R$_{3}$) of the real and mixing events is plotted in the bottom  of
figure \ref{fig5_Eijmin}, together with Breit-Wigner fits. As one can see that the first peak around 0.088 MeV (very close to 0.092 MeV) with 1192 fm/c width corresponds to the ground state of $^{8}$Be and it somehow depends on the detector correction given by the exponential fit. The second peak around 3.05 MeV and 14.2 fm/c width (independent on the detector correction) corresponds to the first excited state of $^{8}$Be, and also higher energy peaks are visible above 10 MeV.

\begin{center}
\includegraphics[scale=0.35]{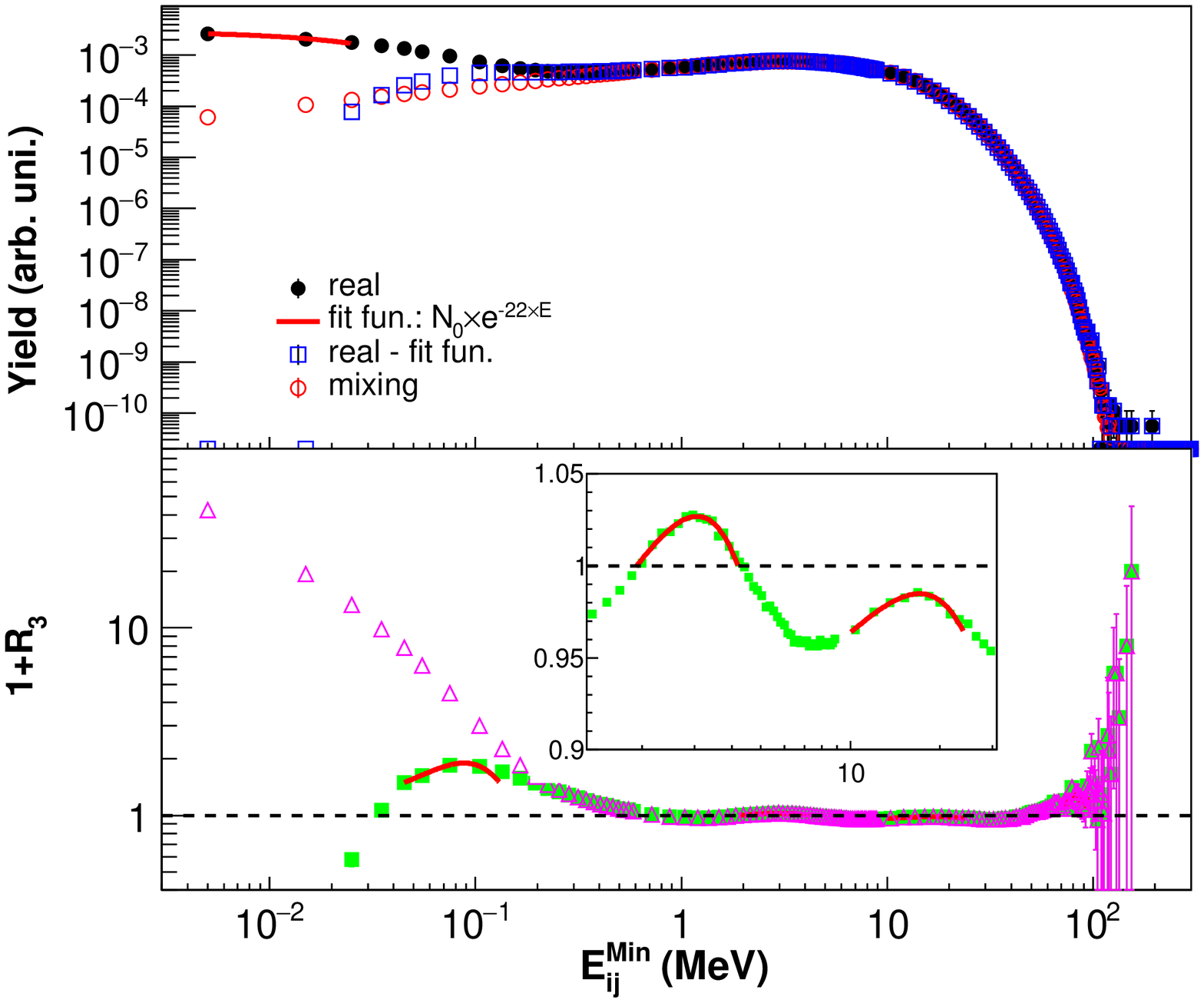}
\figcaption{\label{fig5_Eijmin} (color online) (Top) Relative kinetic energy distribution as a function of the minimum relative kinetic energy. The solid black circles represent data from real events, red open circles are from mixing events, and the blue open squares represent the difference between the real events and the exponential fit function (solid line), which takes into account the experimental error. (Bottom) The ratios of the real data (pink open triangles) and the real data minus the fit function (green solid squares) are divided by the mixing events as a function of the minimum relative kinetic energy. The solid lines are Breit-Wigner fits.
}
\end{center}

In order to determine if we have events with equal relative kinetic energies, we have selected 3$\alpha$ events with  $E_{ij}^{Min.} = E_{ij}^{Mid.} = 0.092\pm\frac{\delta E}{3}$ MeV and decreased the value of $\delta E$ to the smallest value allowed by statistics. In Fig. \ref{fig6_ESEk}, we plot the results for the real (solid black circles) and the mixing (red open circles) events in the upper panels, and their ratio (1+R$_3$) in the bottom panels. Even though the number of real events decreases to almost 90 when the $\delta$E=0.06 MeV case, we can see a hint of a signal around ($E_{ij}^{Lar.} + E_{ij}^{Mid.} + E_{ij}^{Min.} )\times \frac{2}{3}$ - Q $\leq$ 7.47 MeV which is consistent with the suggested Efimov (Thomas) state \cite{HuaPLB2018,SuyalatuarXiv2018,HuaarXiv2018} at an excitation energy of $^{12}$C of about 7.458 MeV.

Similar to Fig. \ref{fig6_ESEk}, we have selected 3$\alpha$ events with  $E_{ij}^{Min.} = 0.092\pm\frac{\delta E}{3}$ MeV, $E_{ij}^{Mid.} = 0.092\times2\pm\frac{\delta E}{3}$ MeV in Fig. \ref{fig7_123Ek}. We can also observe the events where the largest relative energy is three times of the minimum one around ($E_{ij}^{Lar.} + E_{ij}^{Mid.} + E_{ij}^{Min.} )\times \frac{2}{3}$ - Q $=$ 7.64 MeV with different $\delta E$. These events suggest that there are events where the 3$\alpha$ relative energies are in the ratio of 1:2:3.

In Figs. \ref{fig6_ESEk} and \ref{fig7_123Ek}, we can see significant signals around E$^{*}$=7.65 MeV, which is consistent with the famous 0$^{+}$ Hoyle state of $^{12}$C predicted by Fred Hoyle in 1953 \cite{FHoyle1954}.

\begin{center}
\includegraphics[scale=0.45]{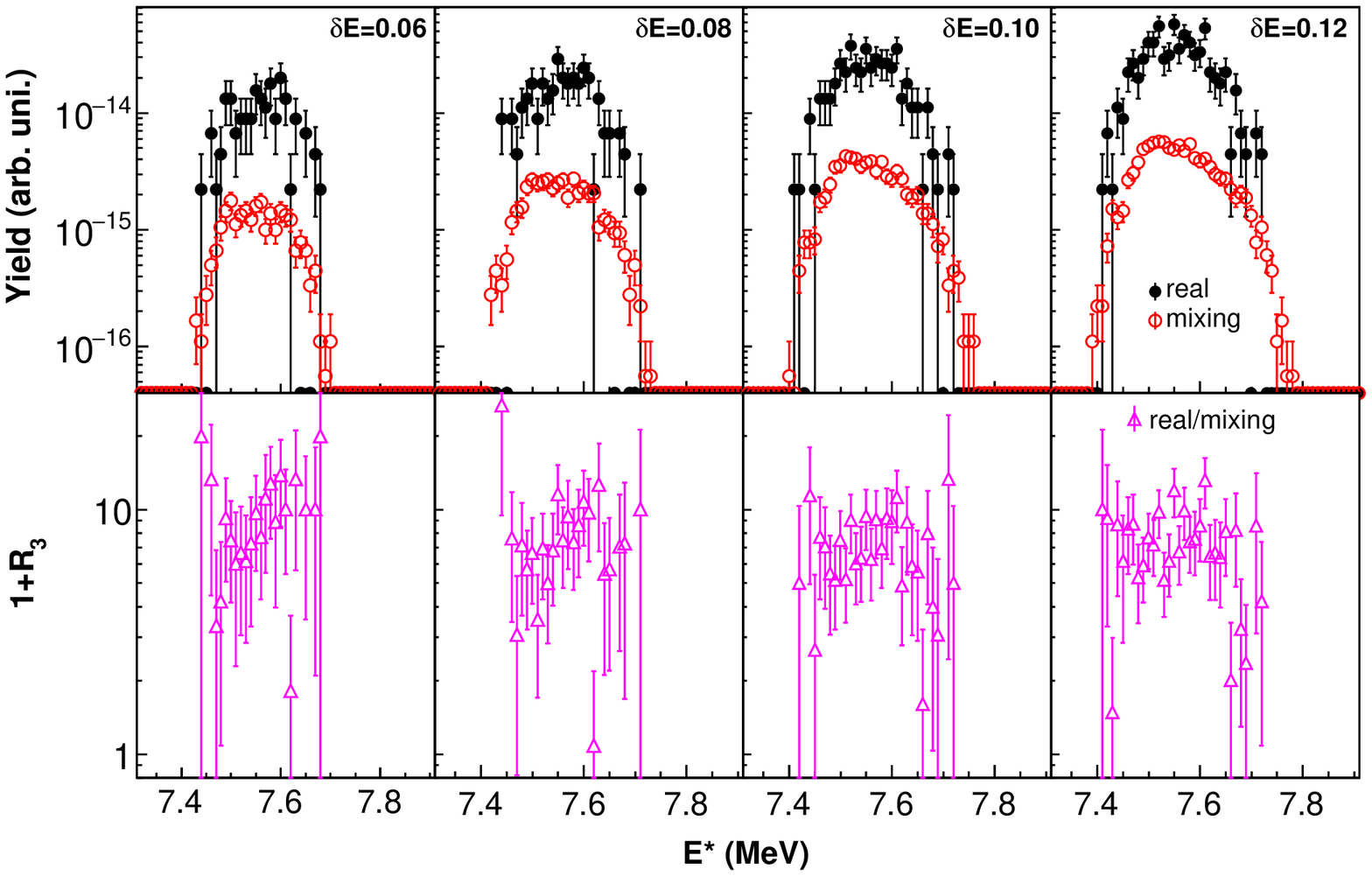}
\figcaption{\label{fig6_ESEk} (color online) The reconstructed excitation energy distributions of $^{12}$C from 3$\alpha$s with $E_{ij}^{Min.} = E_{ij}^{Mid.} = 0.092\pm\frac{\delta E}{3}$ MeV. The solid black circles are from the real events, red open circles are the mixing events, pink open triangles indicate the ratios of the real events to the mixing events.}
\end{center}

\begin{center}
\includegraphics[scale=0.45]{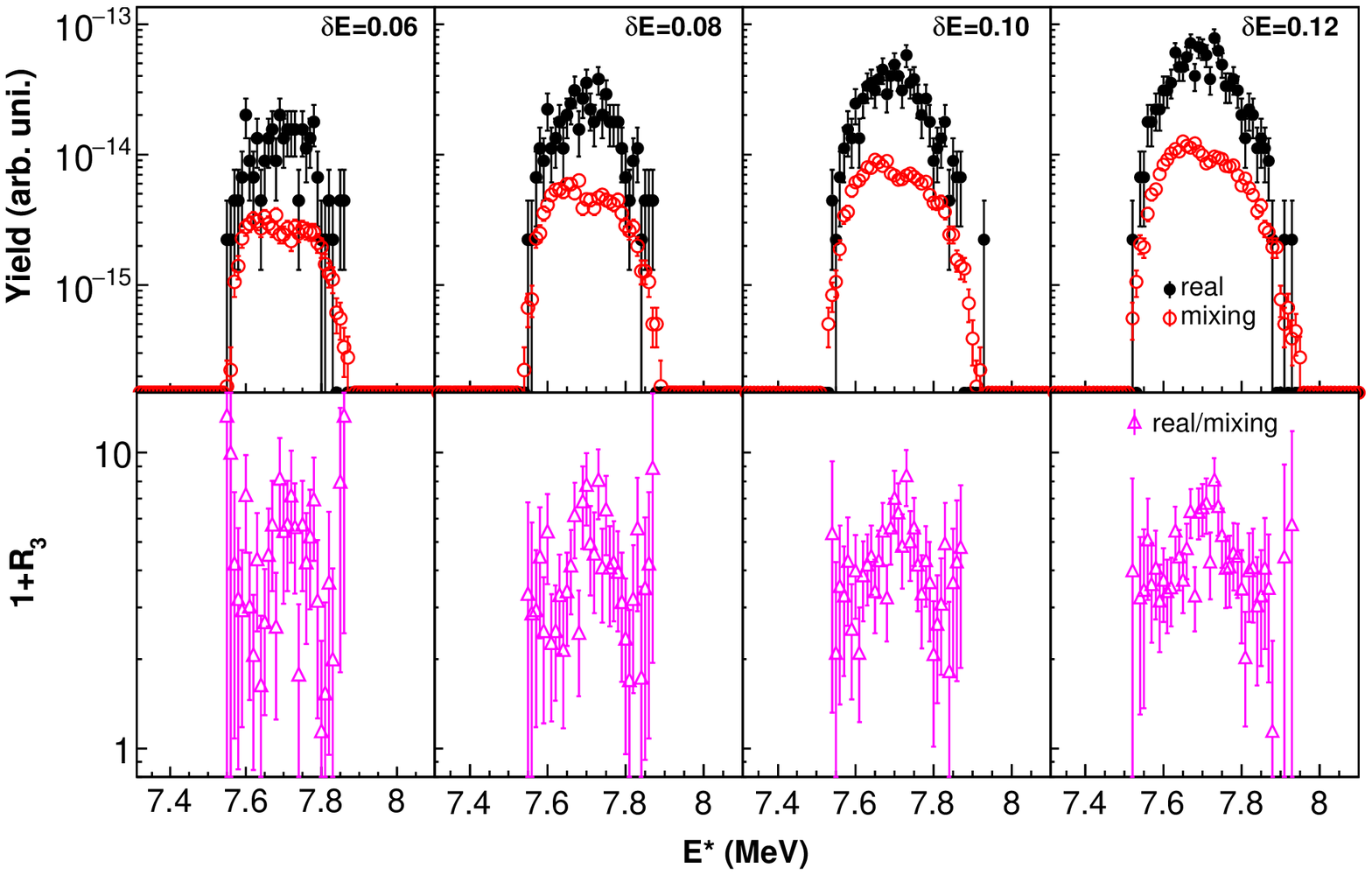}
\figcaption{\label{fig7_123Ek} (color online) The reconstructed excitation energy distributions of $^{12}$C from 3$\alpha$s with $E_{ij}^{Min.} = 0.092\pm\frac{\delta E}{3}$ MeV, $E_{ij}^{Mid.} = 0.092\times2\pm\frac{\delta E}{3}$ MeV. The solid black circles and the red open circles respectively denote the real events, the mixing events, pink open triangles indicate the ratios of the real events to the mixing events. }
\end{center}

\section{Summary}
We have discussed the Efimov (Thomas) states in excited $^{12}$C nuclei in the reactions $^{70(64)}$Zn($^{64}$Ni) + $^{70(64)}$Zn($^{64}$Ni) at beam energy of E/A=35 MeV/nucleon. In order to investigate the $^{12}$C, we analyzed the events with $\alpha$ multiplicity equal to three. The excitation energies of $^{12}$C are reconstructed by considering the three $\alpha$ relative kinetic energies. The interaction between any two of the three $\alpha$ particles provides events with one, two or three $^{8}$Be interfering levels. The events with three relative kinetic energies equal to the ground state energy of $^{8}$Be are found when decreasing the acceptance width. It might be a signature of the Efimov (Thomas) states in $^{12}$C excited energy level of 7.458 MeV. Dedicated experiments with better experimental resolution are suggested in order to exclude any possible experimental effect in the data analysis.

\end{multicols}

\vspace{-1mm}
\centerline{\rule{80mm}{0.1pt}}
\vspace{2mm}

\begin{multicols}{2}

\end{multicols}

\clearpage
\end{CJK*}
\end{document}